\documentclass[reprint,
 amsmath,amssymb,
 aps,
]{revtex4-2}

\usepackage{graphicx}
\usepackage{dcolumn}
\usepackage{bm}
\usepackage{hyperref}
\usepackage[dvipsnames]{xcolor}

\begin{document}

\title{Proposal for a scalable charging-energy-protected topological qubit in a quantum spin Hall system}

\author{Dmitry I. Pikulin}
\affiliation{Microsoft Quantum, Redmond, WA 98052, USA}
\affiliation{Microsoft Quantum, Microsoft Station Q, University of California, Santa Barbara, California 93106-6105, USA}

\date{\today}

\begin{abstract}
The quantum spin Hall edge is predicted to reliably produce Majorana zero modes on the border between magnetic insulator- and superconductor-proximitized regions of the edge.
The direction of magnetization determines the size of the induced magnetic gap and can control the resulting tunnel barrier.
Here we propose a way to avoid magnetic manipulations of the material and use electric-only local control of the barrier.
We follow with a design of a charging-energy-protected qubit and a layout of a quantum computer based on the quantum spin Hall effect.
We estimate relevant scales and show that they allow for testing of these ideas in the near future.
\end{abstract}
\maketitle

\textcolor{Periwinkle}{Introduction ---}
Topological quantum computation is, by design, exponentially protected against local perturbations.
Topological quantum computation is based on the presence of non-Abelian anyons producing ground state degeneracy encoded in non-local degrees of freedom.
This ground state subspace includes the computational subspace of the topological quantum computer and is manipulated either via adiabatic movement or measurement of different combinations of the anyons\cite{nayak2008non}.

Recently the Majorana zero mode (MZM)-based designs for the topological quantum computers have received considerable attention not the least due to experimental progress in observation and manipulation of MZMs\cite{mourik2012signatures, alicea2012new, beenakker2013search, lutchyn2018majorana}.
Signatures of the MZMs in transport have been widely reported and disputed\cite{mourik2012signatures, liu2012zero, kells2012near, nadj2014observation, nichele2017scaling, liu2017andreev, chen2019ubiquitous}.
First studies reporting simple manipulation of MZM-like bound states have also been published, including $4\pi$ Josephson effect\cite{rokhinson2012fractional, wiedenmann20164, laroche2019observation} and $1e$ superconducting charge qubit\cite{van2020photon}.
The studies, though showing the predicted signatures of the MZMs, are not direct evidence of non-Abelian nature of MZMs.
This leaves them open for alternative interpretations.
Successful observation and manipulation of a truly topological qubit is imperative to solidify the MZM claim to be the platform of choice for quantum computing.

The reasons for the difficulty of the decisive measurement of the topological qubit lie in the complications of both manipulation of Majorana zero modes and generation of them in the same material.
The main candidate systems to obtain MZMs include semiconductor-superconductor hybrids\cite{lutchyn2010majorana, oreg2010helical, mourik2012signatures, lutchyn2018majorana}, heavy atom chains on top of a superconductor\cite{nadj2013proposal, nadj2014observation}, and topological insulator (TI)-superconductor composites\cite{elliott2015colloquium, aguado2017majorana}.
Let us go through the pros and cons of the systems.

\begin{itemize}
	\item Heavy atom chains are great for observation of MZMs -- by carefully choosing the material combinations the MZMs can be obtained reliably and can be measured using scanning tunneling microscopy.
	Manipulation of such systems is extremely difficult with current techniques and which makes the platform impractical for quantum computation as of now.
	\item Semiconductor-superconductor hybrids are difficult to tune into the MZM regime. For a practical quantum computer, a large number of the wires should be tuned into the topological regime, which presents the biggest obstacle on the current path of the system to practicality.
	Possibly, careful choice of materials and improvement in quality of the current materials will help solve this problem.
	The advantage of the semiconductor-based design is in their high tunability and well-known control and readout approaches.
	This approach encompasses 2DEG and selective area growth(SAG)-based semiconductors.
	\item TI-based proposals for MZMs have the advantage of the topological regime not requiring tuning.
	However, the manipulation of the MZMs is difficult as magnetic barriers are needed for localization of MZMs. The barriers should also be tunable to allow coupling and de-coupling MZMs.
	There are proposals for 3D TIs\cite{manousakis2017majorana} and 2D TIs without exponential protection\cite{mi2013proposal}.
	In the present work we make a suggestion how to resolve the remaining difficulty and make the 2D TI-based proposal combine the advantages of the guaranteed MZMs and easy manipulations of them.
\end{itemize}

We also note the importance of the charging energy protection for the Majorana qubits. It is predicted that charging energy increase protection against quasiparticle poisoning dramatically compared to the grounded devices\cite{hyart2013flux, mi2013proposal,aasen2016milestones, plugge2017majorana, karzig2017scalable, karzig2020quasiparticle}.

This manuscript is organized as follows: first, we propose a way to make an \textit{electrically tunable magnetic barrier} in a quantum spin Hall (QSH) edge.
Next we discuss how to design a basic element of the MZM-based topological quantum computer, a qubit.
Finally, we discuss the peculiarities of the charging energy-protected qubits in QSH-based designs on large scale, making our proposal scalable and usable in future topological quantum computers.

\textcolor{Periwinkle}{Electrically controlled tunnel barrier ---}
There are a number of ways to form a tunnel barrier in the quantum spin Hall edge proposed. Due to the time-reversal symmetry, in absence of magnetic elements, backscattering is prohibited in the QSH edge. 

Even in presence of time-reversal symmetry breaking, coherent backscattering has been hard to induce in the QSH edge. HgTe and InAs/GaSb, QSH candidates, have been sometimes resistant to the external field\cite{konig2007quantum, du2015robust}. This may be due to the burial of the Dirac point, where the backscattering is most effective, in the bulk band of the semiconductor\cite{skolasinski2018robust}. This effect is present in many QSH materials \cite{murakami2006quantum, qian2014quantum}, but not in WTe$_2$\cite{fei2017edge, wu2018observation} and some HgTe quantum wells\cite{piatrusha2019topological}. There have been additional ways proposed to cause backscattering in the QSH edge, including proximitizing the edge with a magnetic insulator \cite{fu2009josephson}, and gating a region of the material into bulk conductance regime in presence of external perpendicular magnetic field \cite{mi2013proposal}. Only the latter approach allows for electrical control of the tunnel barrier. Making the magnetic control local is an extraordinary challenge for a large enough device as it requires local electromagnets. The proposal~\cite{mi2013proposal} allows for electric control of the tunnel barrier, but it does not allow exponential suppression of conductance and also has effectively small gap protecting the system from quasiparticle poisoning.

\begin{figure}
    \centering
    \includegraphics[width=\linewidth]{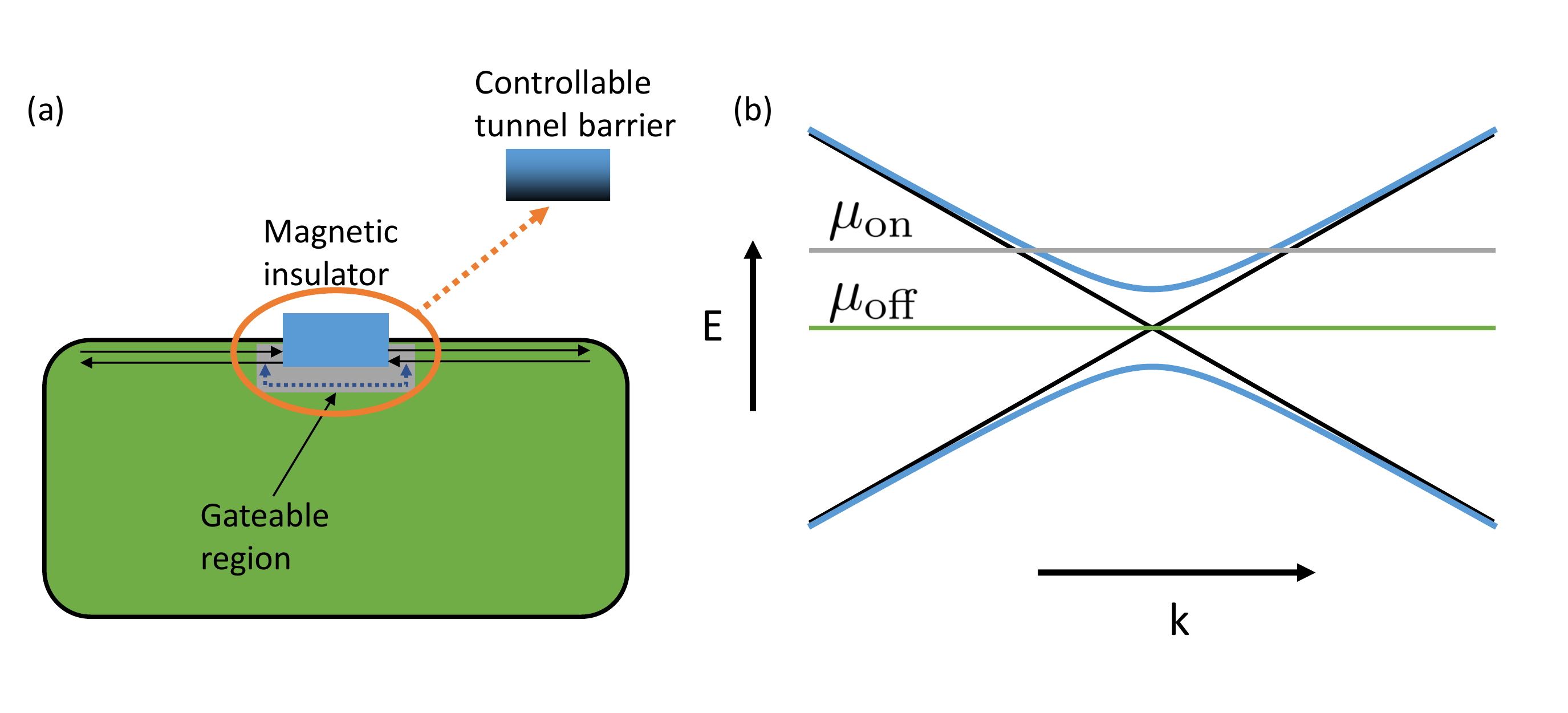}
    \caption{(a) Proposal for controllable tunnel barrier. The magnetic insulator (blue) is in contact with the quantum spin Hall edge, preventing transport inside the induced magnetic gap. The local gate (grey) controls the chemical potential under the magnetic insulator.  We recover transport through the barrier when the chemical potential is tuned to lie outside of the induced gap ($\mu_{\rm on}$). (b) Schematic of the dispersion of the quantum spin Hall edge for the uncovered (black line) and magnetic insulator-covered (blue line) regions. Grey and green lines illustrate the chemical potential positions for the transport "on" and "off" correspondingly.}
    \label{fig:tunnel_barrier}
\end{figure}
In the present work we propose a way to electrically tune the tunnel barrier in the QSH system for localization of a Majorana zero mode \cite{fu2009josephson} and control of electrical circuits based on the QSH\cite{pikulin2019quantum}. The main idea is the following: we would like to use a local magnetic insulator to gap the QSH edge. For the magnetic insulator to be effective we need to locally tune the edge chemical potential to be close to the edge Dirac point. This is achievable by a local gate near the magnetic insulator region. When the gate is tuned to near the Dirac point in the edge spectrum, the magnetic insulator opens a gap near the Fermi energy and hence exponentially suppresses the tunneling across the barrier. However, by tuning the local gate away from the opened gap, we can restore perfect transmission through the barrier. For intermediate gate settings we can obtain any desired transmission, which will be necessary for the topological quantum computation applications we discuss below. The schematics of the barrier and the dispersion relation of the QSH edge near the tunable tunnel barrier are shown in Fig. \ref{fig:tunnel_barrier}a and b correspondingly. 

We note that this may sound as the requirement for our proposal to work is for the gap opened in the edge state to be smaller than the bulk topological gap.
This is not the case, as if the magnetic gap is larger than the topological gap, the local gate can tune the system under it into the TI bulk conduction regime. This still facilitates the transport across the tunnel barrier.

The advantage of our proposal is that it combines the strong sides of what was discussed so far: it allows for electric-only control of the tunnel barrier, and it allows for exponential suppression of the tunneling across the barrier when the Majorana it separates from the rest of the system must stay decoupled.

\textcolor{Periwinkle}{Parameters estimation ---}
Let us now estimate the parameters of the tunnel barrier needed to efficiently decouple the Majorana zero modes from the environment for the relevant material combinations. We take EuS as the magnetic insulator and edge state velocities of HgTe, InAs/GaSb, and WTe$_2$.

The spin splitting of the bands in EuS is large, $\approx 0.36$eV \cite{hao1990spin}. However, let us restrict the induced gap to the conservative estimate of the induced exchange constant of $\Delta_f\approx 3.5$meV. This gives a wavefunction decay length of $\xi_f = \frac{v_F}{\Delta_f}$. $v_F$ is $\approx 0.37$eV~nm for HgTe\cite{bernevig2006quantum}, $\approx 0.072$eV~nm for InAs/GaSb\cite{liu2008quantum}, and $\approx 0.2$eV~nm for WTe$_2$\cite{pulkin2020controlling, pulking2020private}. In turn the decay length under the barrier of $\xi_f\approx 100$nm for HgTe, $\xi_f\approx 20$nm for InAs/GaSb, and $\xi_f\approx 50$nm for WTe$_2$. The rate of tunneling from the Majorana across the tunnel barrier can be estimated as:
\begin{align}
\Gamma \approx \frac{1}{\pi} g \Delta,
\end{align}
where $g$ is the dimensionless conductance through the barrier, and $\Delta$ is the topological gap, i.e. induced superconducting gap in our case.
For the optimal case of tuning to the middle of the barrier to minimize the tunneling:
\begin{align}
	g=\frac{1}{\cosh^2 (L/\xi_f)}\approx 4 e^{-2 L/\xi_f},
\end{align}
for $L\gg \xi_f$.
If we take the induced gap to be $2$K, $\Gamma \approx g \times 8 \times 10^{10}$Hz.
Thus, to achieve the time to leak across the barrier of $\sim 100\mu$s, tunnel barriers of lengths $1000$nm, $200$nm, and $500$nm respectively are sufficient.
These tunneling times directly convert to the coherence times of the qubits introduced below.
We summarize the parameters in Table~\ref{tbl:parameters}.
Based on these numbers, InAs/GaSb in theory seems to be the most favorable material, where barrier just slightly longer than $200$nm would be sufficient for any long-term quantum computation application \footnote{It is important to note that although there was some success in measuring the edge modes, convincing demonstration of the bulk gap together with topologically protected conductance is still missing\cite{knez2011evidence, nichele2016edge, han2019anomalous}.}
We also note the danger of the longer barriers due to the possible disorder in the system -- disorder may induce states inside the barrier and significantly suppress coherence times of the qubit.

\textcolor{Periwinkle}{Qubit designs ---}
\begin{table}
	\begin{tabular}{ c | c | c }
		Material & $v_f$ & $L_{\rm barrier}$ \\ \hline
		HgTe & $0.37$eV nm & $\approx 1\mu$m \\
		InAs/GaSb & $0.072$eV nm & $\approx 0.2\mu$m\\
		WTe$_2$ & $0.2$eV nm & $\approx 0.5\mu$m
	\end{tabular}
	\caption{Parameters of the common QSH materials. $L_{\rm barrier}$ is such that  qubit coherence time is predicted to exceed 100$\mu$s}
	\label{tbl:parameters}
\end{table}
\begin{figure}
    \centering
    \includegraphics[width=\linewidth]{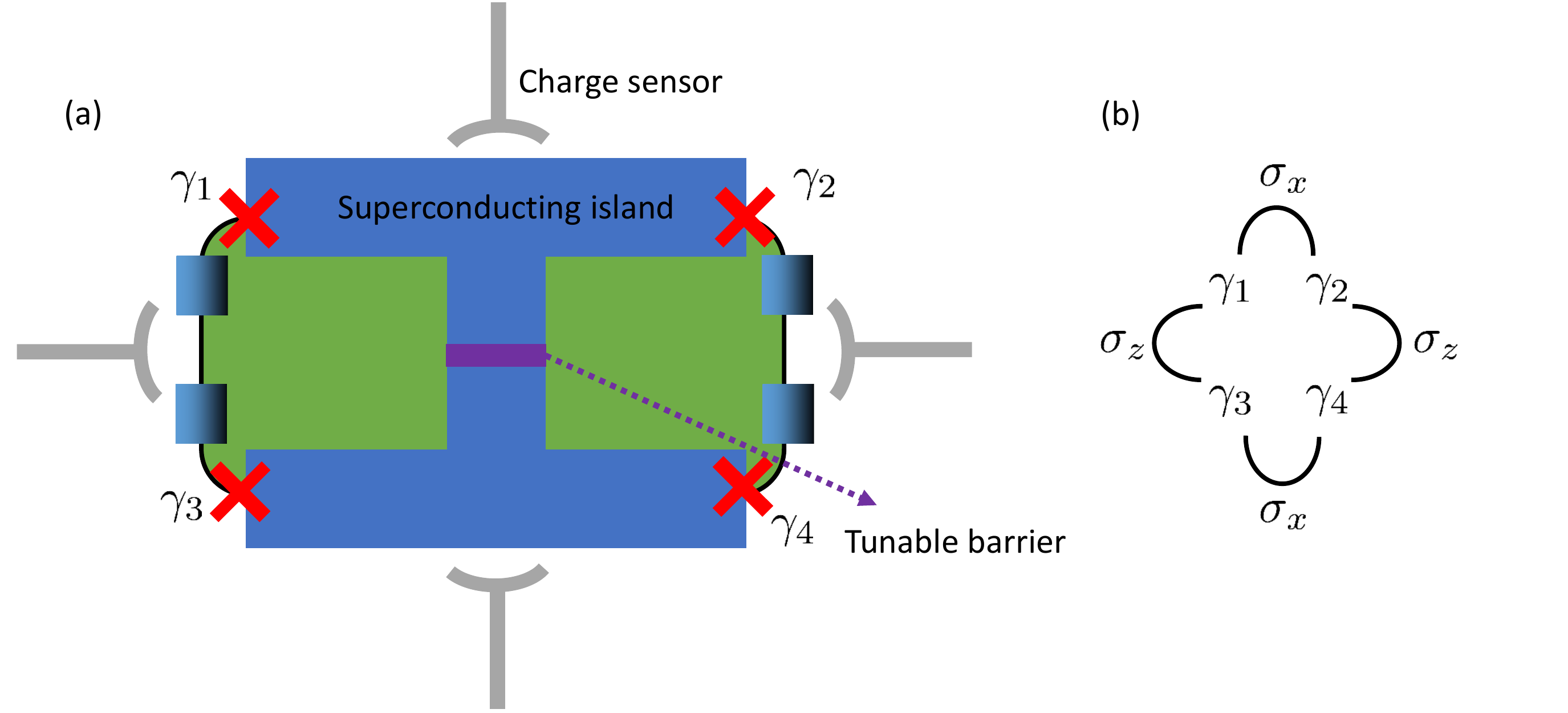}
    \caption{(a) Proposal for design of a single topological qubit with topological measurements in $\sigma_x$ and $\sigma_z$ bases. The design include two superconducting islands (blue) connected via a tunable tunnel barrier (purple) and two edges of the QSH system (green) intercepted by two tunable magnetic barriers each. Operations include opening and closing the tunnel barriers and measuring charge and quantum capacitance using the charge sensors (grey) (b) Majorana fusion operations performed by the measurements in (a). Topological basis of the fusion operation should make the occupied/empty result in the second measurement of $\sigma_x$-$\sigma_z$ cycle very close to $50\%$.}
    \label{fig:one_qubit}
\end{figure}
The single-qubit design we propose in this section has the advantage of being very versatile and allowing for two methods of the qubit readout. The choice between the two can be made depending on what is experimentally more challenging -- to make a quantum dot measurement or to make a tunable Josephson junction between two superconducting islands in the QSH bulk. Either design allows one to measure the MZM fusion.

The proposal is illustrated in Fig. \ref{fig:one_qubit}. The qubit consists of four MZMs $\gamma_i$ localized between the superconducting region and the ferromagnetic insulator tunable barrier. Ground states of the definite parity form the qubit subspace $|0, 0\rangle$ and $|1, 1\rangle$ in the basis of parity states of $i\gamma_1 \gamma_2$ and $i \gamma_3 \gamma_4$.
The design includes a new element, tunable tunnel barrier between superconducting islands via QSH bulk.
This is achieved by putting a gate near the junction region.
In the closed regime the gate is inactive and to open the tunnel barrier we suggest to locally tune the QSH system into bulk conducting regime.
Opening and closing the tunnel barrier controls mutual charging energy of the upper and lower superconducting islands in Fig.~\ref{fig:one_qubit}a.
Measurements of different pairs of Majoranas can be performed as follows: 
\begin{itemize}
    \item $i\gamma_1 \gamma_2$ and $i \gamma_3 \gamma_4$ can be measured by cutting the connection between the two superconducting islands. This introduces separate charging energies for the two islands and allows the measurement of their charge. These two measurements are equivalent when the system is within the qubit subspace, but can be performed simultaneously to check for some of the errors getting the system out of the qubit subspace.
    \item $i\gamma_1 \gamma_3$ and $i \gamma_2 \gamma_4$ can be measured using the quantum dot readout \cite{karzig2017scalable}. Quantum capacitance of the quantum dot coupled to a pair of Majoranas depends on the parity of the pair, thus providing the necessary measurement.
\end{itemize}

This design allows to perform two important measurements for the demonstration of topological qubit, coherent oscillations and Majorana fusion. Coherent oscillations can be observed if direct $i \gamma_1 \gamma_3$ coupling is turned on for a small time and then turned off, after which the $i \gamma_1 \gamma_2$ parity is measured. Such operation provides $\sigma_x$ rotation and measurement in $\sigma_z$ basis. Majorana fusion can be performed if, first, measurement of $i\gamma_1 \gamma_2$ is performed. This initializes the qubit in the $\sigma_z$ eigenstate. In the language of non-abelian anyons this is equivalent to having only $\gamma_3$ and $\gamma_4$ in the low-energy subspace. When the measurement is done, one is creating $\gamma_1$ and $\gamma_2$, after which the measurement of $i \gamma_1 \gamma_3$ gives fusion result of one new MZM and one existing before. This should give $50\%$ chance of $0$ or $1$ result, but perfectly correlated with the measurement of $i \gamma_2 \gamma_4$. Such measurement would demonstrate the fusion rule of Majorana zero modes and hence their non-abelian statistics.

The design also allows for a transmon measurement instead of charge sensing. The top and bottom superconductors can be connected to the transmission lines and the spectrum of the qubit can be measured\cite{ginossar2014microwave}. We note that the design allows for separate control of the Majorana coupling through the QSH edge and the Josephson coupling through the tunable barrier. This should produce clean signatures of the 1e transmon predicted in\cite{ginossar2014microwave}.

Finally we note that the best visibility of the quantum dot measurement is achieved when the level spacing in the quantum dot is much larger than the tunnel coupling across the tunnel barriers.
It is highly preferable that the level spacing of the dot is larger than temperature as in the opposite regime occupation of the higher levels start to play an important role and the visibility of the measurement is severely limited.
Typical electron temperatures in experiment are $\approx 5\mu$eV, thus limiting the length of the quantum dots in the figures presented to less than $\approx14\mu$m for InAs/GaSb and even larger for the other materials.
Thus this seems to be a weak bound on the level spacing.

\textcolor{Periwinkle}{Scalable design ---}
\begin{figure}
	\centering
	\includegraphics[width=\linewidth]{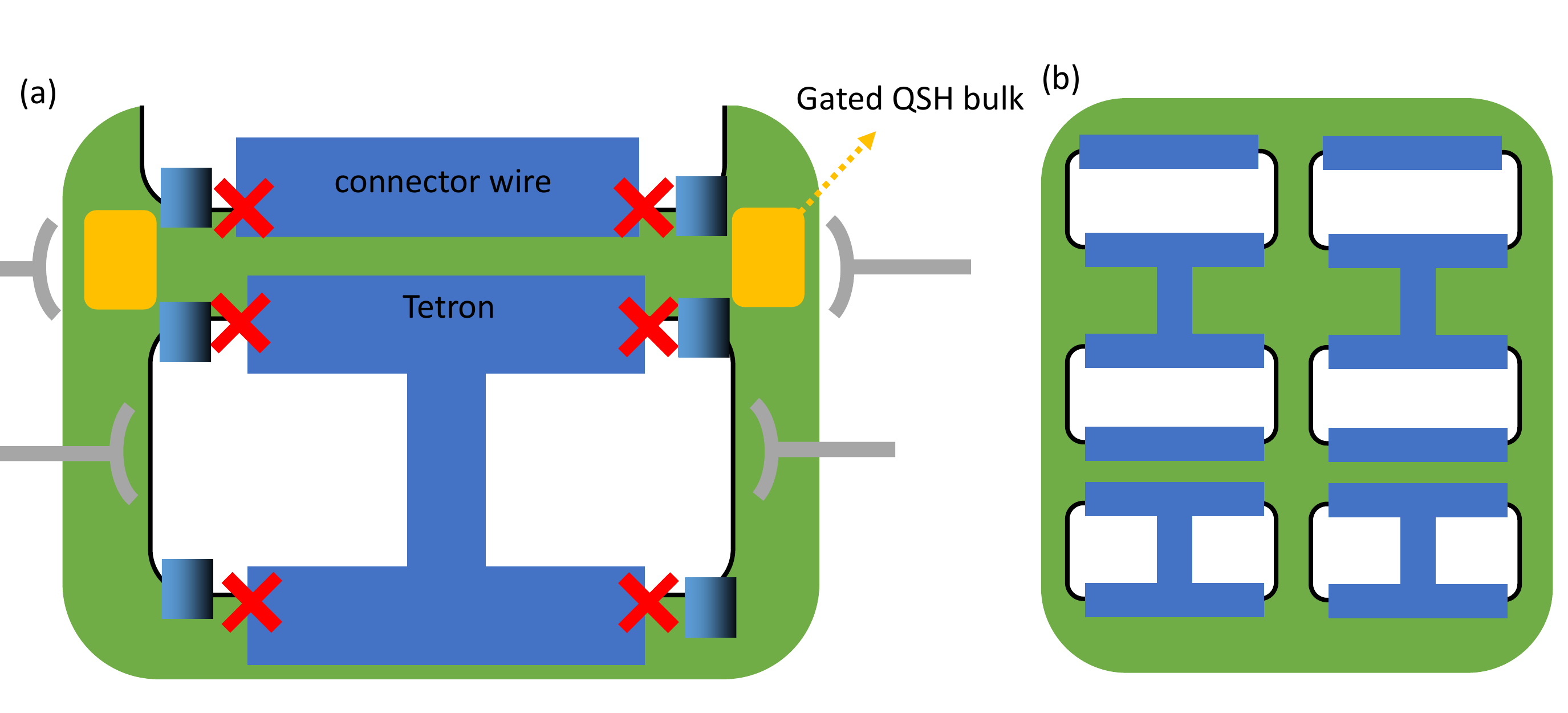}
	\caption{(a) Proposal for design of a single topological qubit with topological measurements in $\sigma_x$ and $\sigma_z$ bases based on quantum dot readout only for use in a scalable pattern. Different edges of the QSH system are connectable by tunable quantum dots (yellow) (b) Example of a layout of the tetron qubits that allows for all in-qubit operations and as well as entangling ones between nearby qubits on a rectangular grid.}
	\label{fig:tetron_based}
\end{figure}
We now proceed with the design that allows coupling of multiple topological qubits together and performing the measurment-based charging-energy-protected topological quantum computation. The requirements for such a design include: \textit{(i)} multiple topological qubits with all possible measurements; \textit{(ii)} two-qubit measurements forming a graph connecting all the qubits; \textit{(iii)} possibility to perform non-universal $\pi/8$ gate. As we will show the QSH system allows to satisfy all the requirements.

The design in the previous section did not allow $\sigma_y$ measurement and had a non-uniform way to measure the $\sigma_x$ and $\sigma_z$. While suitable for an early qubit prototype, this drawbacks can be fixed with extra design knobs. Those are based on~\cite{karzig2017scalable}, and we suggest two possible design choices for the QSH system: tetron-based design and hexon-based design. 

Tetron-based designs are shown in Fig. \ref{fig:tetron_based}. Fig. \ref{fig:tetron_based}a shows a single repeating element of the scalable design. By employing the quantum dots to the left and right of the tetron qubit one measures the left and right pairs of MZMs. To connect the decoupled edges of the QSH system we propose adding gate-controlable regions of the QSH system which can be tuned into the bulk conductance regime. Using regions to connect a wire with two MZMs on the side, one can connect the right and left side of the tetron qubit and measure pairs of MZMs with elements from both sides.

In Fig. \ref{fig:tetron_based}b we show a possible layout of a scalable setup -- one can continue the proposed layout translationally invariant both horizontally and vertically. As in Fig. \ref{fig:tetron_based}a, any measurement on a given tetron is possible, plus an entangling one between nearby tetrons is straightforward to make using the intervening quantum dots in between the qubits. One entangling measurement is enough for a unversal set of gates. We also note that this can trivially be extended to the hexon qubits~\cite{karzig2017scalable, pikulin2019quantum}.

As shown in ref.~\cite{khindanov2020visibility}, the two-qubit measurement is best performed when both quantum dot participating are at resonance.
In such a regime the constraints on the coupling to the quantum dot and the level spacing of the dots are the same as in a single-qubit measurement, thus the dots are still weakly constrained for the two-qubit measurement as well.

\textcolor{Periwinkle}{Outlook and conclusions ---}
In this manuscript we used the quantum spin Hall systems as an example system to realize the proposals, however that is not the only system where the proposed architecture can be realized. Another possible platform is a three-dimensional second order topological insulator, which has topologically protected hinge states. Those are direct analogues of the QSH edge states. Thus, a 3D 2nd order TI system where the top surface is curved in the same shape as the QSH planes in our proposal would work just as well, pending the gap can be efficiently opened in the hinge state\cite{queiroz2019splitting}.

We note that the controllable Josephson junctions shown in Fig.~\ref{fig:one_qubit} allow for coupling and decoupling islands containing only pairs of Majoranas.
This allows to realize analogues of design for fermionic quantum computation using Majoranas in our system\cite{o2018majorana}.

In summary, we have shown the way to create an electrically tunable tunnel barrier in a quantum spin Hall edge for which the transparency can be exponentially controlled.
We have also presented a way to use the design to build a single- or many-qubit devices for uses in topological quantum computing.
We finally note that the electrical control of the barrier can help facilitate the long-sought after direct observation of the Majorana zero mode in a QSH edge.

We thanks Aleksei Khindanov, Christina Knapp, Torsten Karzig, and Roman Lutchyn for useful discussions and comments.

\bibliography{bibliography}

\end{document}